\documentclass[journal=jacsat,manuscript=article]{achemso}

\usepackage{chemformula} 
\usepackage[T1]{fontenc} 
\usepackage{siunitx}
\usepackage[defaultcolor=red, final]{changes}

\author{Yubin Huang}
\affiliation[Université catholique de Louvain (UCLouvain)]
{Institute of Condensed Matter and Nanosciences, Université catholique de Louvain (UCLouvain), 1348 Louvain-la-Neuve, Belgium}
\author{Jean Spiece}
\affiliation[Université catholique de Louvain (UCLouvain)]
{Institute of Condensed Matter and Nanosciences, Université catholique de Louvain (UCLouvain), 1348 Louvain-la-Neuve, Belgium}
\author{Tetiana Parker}
\affiliation[Drexel University]
{A. J. Drexel Nanomaterials Institute and Department of Materials Science and Engineering, Drexel University, Philadelphia, PA 19104, USA}
\author{Asaph Lee}
\affiliation[Drexel University]
{A. J. Drexel Nanomaterials Institute and Department of Materials Science and Engineering, Drexel University, Philadelphia, PA 19104, USA}
\author{Yury Gogotsi}
\affiliation[Drexel University]
{A. J. Drexel Nanomaterials Institute and Department of Materials Science and Engineering, Drexel University, Philadelphia, PA 19104, USA}
\author{Pascal Gehring}
\affiliation[Université catholique de Louvain (UCLouvain)]
{Institute of Condensed Matter and Nanosciences, Université catholique de Louvain (UCLouvain), 1348 Louvain-la-Neuve, Belgium}
\email{pascal.gehring@uclouvain.be}

\title
  {Violation of the Wiedemann-Franz law and ultra-low thermal conductivity of Ti$_3$C$_2$T$_x$ MXene}


\begin{document}

\begin{tocentry}
  \centering
  \includegraphics[width=5.5cm]{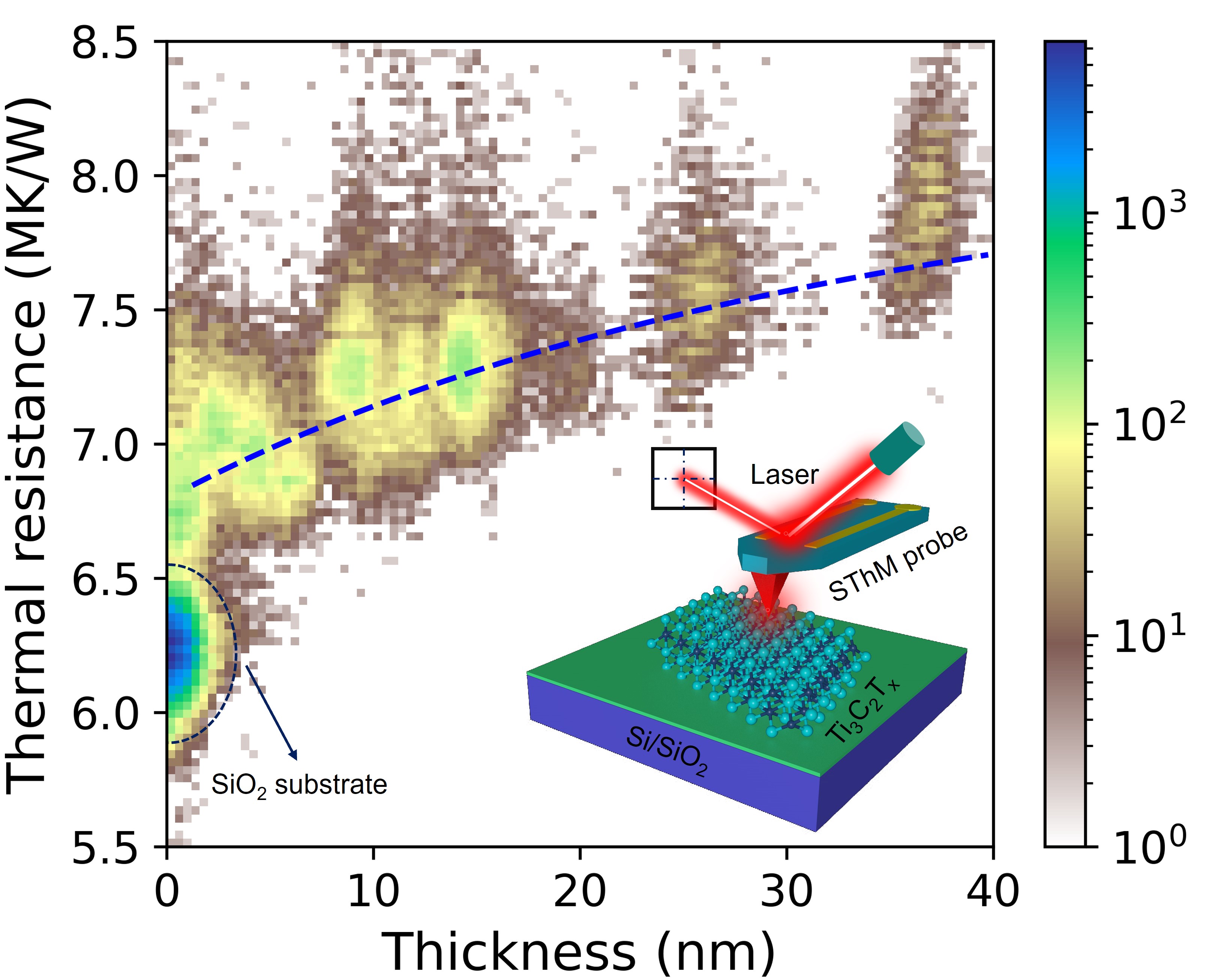}
\end{tocentry}

\begin{abstract}
  The \deleted{remarkably} high electrical conductivity and good chemical stability of MXenes offer hopes for their use in many applications, such as wearable electronics, energy storage, or electromagnetic interference shielding. While their optical, electronic and electrochemical properties have been widely studied, the information on thermal properties of MXenes is scarce. In this study, we investigate the heat transport properties of Ti$_3$C$_2$T$_x$ MXene single flakes using scanning thermal microscopy and find exceptionally low anisotropic thermal conductivities within the Ti$_3$C$_2$T$_x$ flakes, leading to an effective thermal conductivity of 0.78 $\pm$ 0.21 W m$^{-1}$ K$^{-1}$. This observation is in stark contrast to the predictions of the Wiedemann-Franz law, as the estimated Lorenz number is only 0.25 of the classical value. Due to the combination of low thermal conductivity and low emissivity of Ti$_3$C$_2$T$_x$, the heat loss from it is two orders of magnitude smaller than that from common metals. Our study \replaced{explores}{sheds light on} the heat transport mechanisms of MXenes and highlights a promising approach for developing thermal insulation, two-dimensional thermoelectric, or infrared stealth materials.
\end{abstract}

\added{\noindent \textbf{Keywords:} MXene, thermal transport, thermal conductivity, Wiedemann-Franz law, scanning thermal microscopy.}
\section{Introduction}
Materials with low thermal conductivity are of utmost importance for a plethora of applications ranging from thermal insulation, to heat shields, and stealth materials, as well as to thermoelectrics \cite{qian2021phonon}. In this context, 2D materials have gained increasing attention since they can possess very low lattice thermal conductivities, because of their two-dimensional crystal structure and phonon-boundary scattering. For example, WSe$_2$ \cite{chiritescu2007ultralow}, gallium phosphide (GaP) \cite{shen2022two} and indium selenide (InSe) \cite{buckley2021anomalous(inse)} reached low values of 0.048 W m$^{-1}$ K$^{-1}$, 1.52 W m$^{-1}$ K$^{-1}$ and 28.7 W m$^{-1}$ K$^{-1}$, respectively, which were explained by the strong phonon anharmonicity and boundary scattering. However, these materials often have a low electrical conductivity, making them impractical for use in thermoelectric heat engines. This kind of green energy harvester can generate electricity from the heat with an efficiency proportional to the figure of merit $ZT = \frac{\sigma S^2 T}{\kappa}$, where $\sigma$, $S$ and $\kappa$ are the electrical conductivity, the Seebeck coefficient, and the total thermal conductivity, respectively. Therefore, materials with \textit{low} thermal conductivity and \textit{high} electrical conductivity are desired, a property that is inherently difficult to achieve: in most electrically conductive solids, charge carriers also transport heat. Indeed, the electronic thermal conductivity ($\kappa_\mathrm{e}$) is related to the electrical conductivity via the Wiedemann-Franz (WF) law. It states that the ratio of $\kappa_\mathrm{e}$ to $\sigma$ at a given temperature ($T$) is a constant called the Lorenz number ($L_0$): 
\begin{equation}
   L = \frac{\kappa_{e}}{\sigma T} \equiv L_{0} = 2.44 \times 10^{-8}~{\mathrm{W} \Omega \mathrm{K}^{-2}}.  
   \label{eq1}
\end{equation}

In recent years, it has been experimentally demonstrated that this \textit{fundamental} law can be violated in some materials like graphene \cite{crossno2016observation}, VO$_2$ \cite{lee2017anomalously} or in quantum-confined 0D systems \cite{majidi2022quantum}. A key ingredient for this violation appears to be strong correlations of quasiparticles, such as strong electron-phonon coupling \cite{crossno2016observation}. To this end, a large family of 2D materials, transition metal carbides and nitrides (MXenes) \cite{vahidmohammadi2021world}, might be able to demonstrate this violation because some MXenes like Ti$_3$C$_2$T$_x$ have a highly tunable metallic-like conductivity of 5000 to over 20000 S cm$^{-1}$ \cite{mathis2021modified} combined with strong electron-phonon interactions \cite{zhang2022simultaneous, guzelturk2023understanding, colin2023ultrafast}. However, while their electronic properties have been widely investigated, experimental studies on their thermal transport properties remain lacking, especially on the intrinsic thermal conductivity of isolated single flakes. 

In this work, we present local thermal transport measurements on Ti$_3$C$_2$T$_x$ MXene single flakes by using scanning thermal microscopy (SThM) at room temperature. By measuring the thermal conductance variation with varying flake thickness, we extracted an ultra-low thermal conductivity of 0.78 W m$^{-1}$ K$^{-1}$. It coexists with a high electrical conductivity of $>$4430~S cm$^{-1}$ which leads to a strong violation of the WF law in Ti$_3$C$_2$T$_x$ single flakes, with $L$ = 0.25$L_0$. Therefore, our work demonstrates that MXenes provide highly promising 2D materials for ultra-thin heat insulation or thermoelectric applications.

\section{Results and discussion}

\subsection{Synthesis and characterization of Ti$_3$C$_2$T$_x$ flakes}

\begin{figure}[htbp] 
  \centering
  \includegraphics[width=15cm]{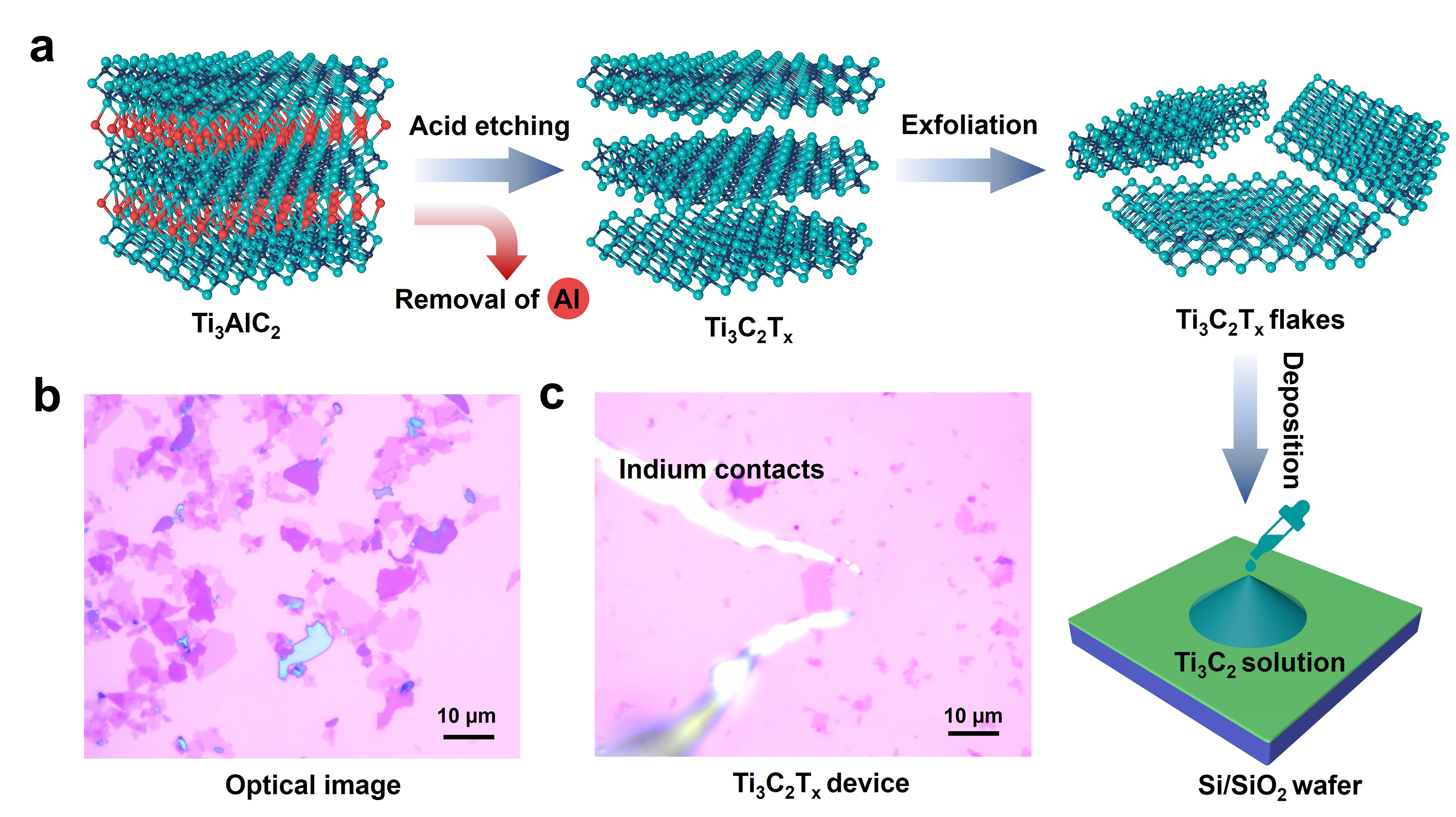} 
  \caption{(a) Schematic of the preparation process of Ti$_3$C$_2$T$_x$ MXene single flakes and devices. (b) Optical image of monolayer and few-layers Ti$_3$C$_2$T$_x$ flakes on SiO$_2$ substrate. (c) A two-terminal Ti$_3$C$_2$T$_x$ device of single flake contacted with two indium needles.}  
  \label{fig:example} 
\end{figure}

Ti$_3$C$_2$T$_x$ MXene flakes were synthesized following a well-established process \cite{ghasali2024biofouling}, as described in the Methods section. Briefly, we first etched the aluminium layers of the MAX phase (Ti$_3$AlC$_2$) in acidic solution, followed by liquid exfoliation to prepare a colloidal suspension of Ti$_3$C$_2$T$_x$ flakes in water (see Figure 1a). A detailed structure characterization of Ti$_3$C$_2$T$_x$ can be found in our previous work \cite{han2020beyond}. Raman spectroscopy was also performed on monolayer Ti$_3$C$_2$T$_x$. We observed a typical peak at around 203 cm$^{-1}$ (see Figure S1, SI), which corresponds to out-of-plane vibration mode $A_\mathrm{1g}$ \cite{sarycheva2022tip}. 

Isolated high-quality Ti$_3$C$_2$T$_x$ flakes with lateral sizes of $\qtyrange{5}{20}{\micro\metre}$ were obtained by drop-casting the dispersion on a Si/SiO$_2$ wafer (see Experimental methods). Figure 1b shows an optical image of a typical deposit. From their optical contrast (in combination with atomic force microscopy (AFM), Figure 2a), the thickness of Ti$_3$C$_2$T$_x$ flakes can be obtained and flakes with thicknesses varying down to the monolayer limit ($\approx$ $\qty{1.8}{\nano\metre}$, Figure 2e) are identified. For electrical conductance measurements, we fabricated devices by nano-scale soldering of indium \cite{girit2007soldering, razeghi2023single} to the single flakes (Figure 1c).

\subsection{Thermal transport in Ti$_3$C$_2$T$_x$ flakes}

To investigate the thermal transport properties of isolated Ti$_3$C$_2$T$_x$ flakes with different thicknesses, from monolayer to tens of layers, we used ambient SThM at room temperature ($T_{room}$ = 293 K). SThM is a contact mode atomic force microscopy technique in which microfabricated probe is equipped with a Pd electrical resistor on the cantilever and close to the tip apex (see Figure 2a). This resistor is used as a temperature sensor and local heater, and its resistance is measured using a custom-made Wheatstone bridge \cite{spiece2019quantitative,spiece2018improving}. Then, the probe is brought into contact with the sample surface. When the mechanical contact occurs, a sharp drop of the probe temperature is observed due to the extra heat transfer opened at the tip apex. Finally, when the tip raster scans the sample surface, the probe temperature is mapped onto the sample topography allowing to calculate the heat flow into the sample and its thermal resistance.

\begin{figure}[htbp] 
  \centering
  \includegraphics[width=15cm]{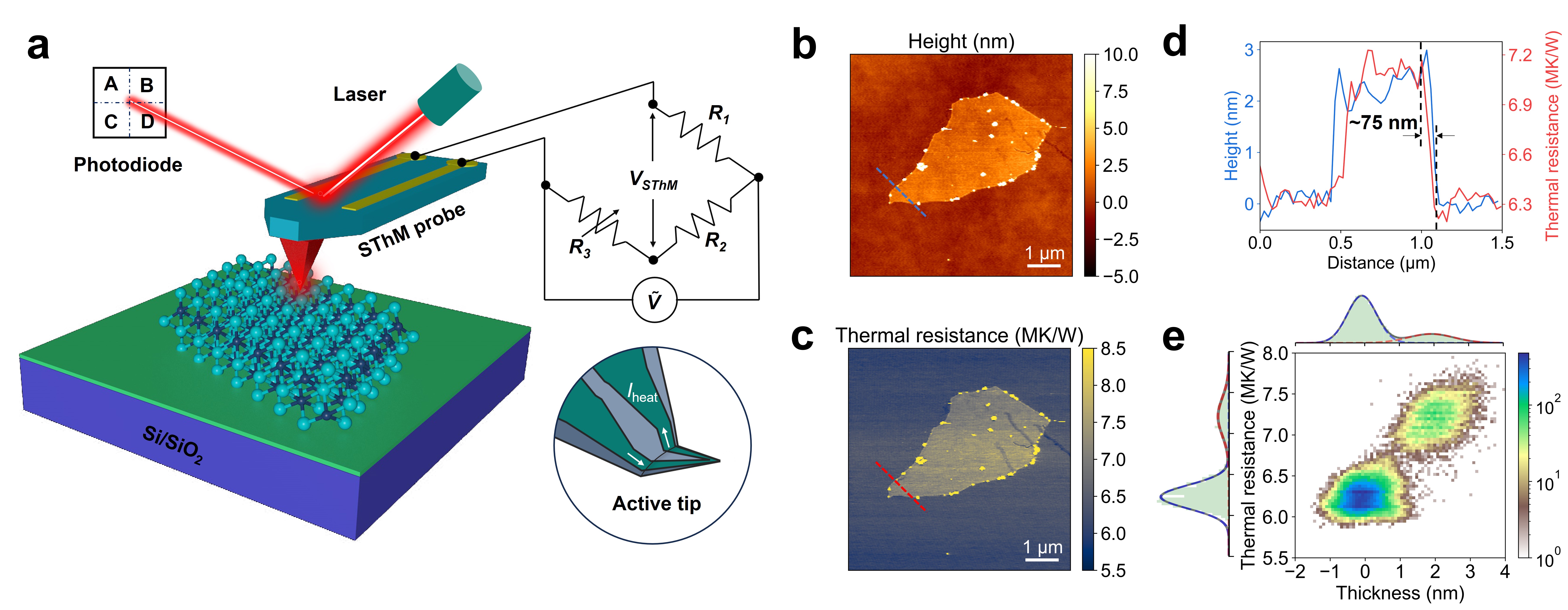} 
  \caption{(a) Schematic of the SThM setup for measuring local thermal transport properties of Ti$_3$C$_2$T$_x$ flakes. The SThM probe can be used for temperature sensing (with a palladium circuit at the tip apex). Thermal transport signal is measured from Wheatstone bridge output. (b) Topography and (c) thermal resistance maps of monolayer Ti$_3$C$_2$T$_x$ on SiO$_2$ acquired in an ambient environment. (d) Height and thermal resistance curves along the dashed blue line in (b) and red line in (c). Vertical dashed lines are used to estimate the lateral thermal resolution and the tip radius of the SThM probe, which are $\qty{38}{\nano\metre}$ and $\qty{75}{\nano\metre}$, respectively (see Note S2, SI). (e) 2D histogram showing the relationship between thickness and thermal resistance of the entire image, mean values can be obtained from the Gaussian fit.} 
  \label{fig:example} 
\end{figure}

Figure 2b and c show a topography and thermal resistance map of a single layer Ti$_3$C$_2$T$_x$ flake simultaneously obtained using SThM (single approach-retraction curves which are used for the conversion between the SThM signal and thermal resistance, and details about this conversion are provided in Note S2, SI). We observed a strong thermal resistance contrast between the Ti$_3$C$_2$T$_x$ monolayer and the SiO$_2$ substrate. The thermal resistance increases from 6.3 MK W$^{-1}$ on bare SiO$_2$ to 7.1 MK W$^{-1}$ on monolayer Ti$_3$C$_2$T$_x$ flakes (see Figure 2d). It is worth to note that while such increased thermal resistance can point towards a lower thermal conductivity of Ti$_3$C$_2$T$_x$ compared to that of SiO$_2$, there can be other factors playing a significant role in the apparent thermal transport: i) the thermal contact between the SThM probe and surface can be different for SiO$_2$ and Ti$_3$C$_2$T$_x$; ii) the newly formed interface between the monolayer and SiO$_2$ substrate can lead to enhanced phonon scattering and thus increased thermal resistance \cite{menges2013thermal,evangeli2019nanoscale}; iii) monolayer flake tends to present a thermal barrier, since it’s enhanced in-plane thermal spreading is poor when compared to the new interface resistance \cite{menges2013thermal}.

To access the thermal \textit{conductivity} of Ti$_3$C$_2$T$_x$, we follow an approach developed in \cite{spiece2021quantifying, spiece2022low, gonzalez2023direct} that reduces the impact of the effects (i-iii) discussed above. To this end, a model (see Experimental Section and SI) is used to fit the thickness-dependence of the thermal resistance. To extract the relationship between thermal resistance and the thickness of Ti$_3$C$_2$T$_x$ flakes, we performed a pixel-to-pixel correlation of height and thermal maps to create 2D histograms (see Figure 2e). This method considers all data points -- and does not rely on arbitrarily chosen line-cuts -- and we process the data statistically, which reduces experimental errors caused by contaminations or artifacts. For the isolated single-layer case, this histogram shows a two-dimensional Gaussian distribution, and we individually fitted the thickness and thermal resistance with Gaussian functions to obtain their mean values and corresponding standard deviations. For example, the monolayer Ti$_3$C$_2$T$_x$ has a thermal resistance of 7.2 $\pm$ 0.23 MK W$^{-1}$. We applied this method to other SThM data of MXene flakes with different thicknesses (see Figures S4, SI) to create a 2D histogram of all data as shown in Figure 3a. With increasing flake thickness, the thermal resistance increases and approaches the thermal resistance of the bulk Ti$_3$C$_2$T$_x$ in the limit of infinite thickness.

\begin{figure}[htbp] 
  \centering
  \includegraphics[width=15cm]{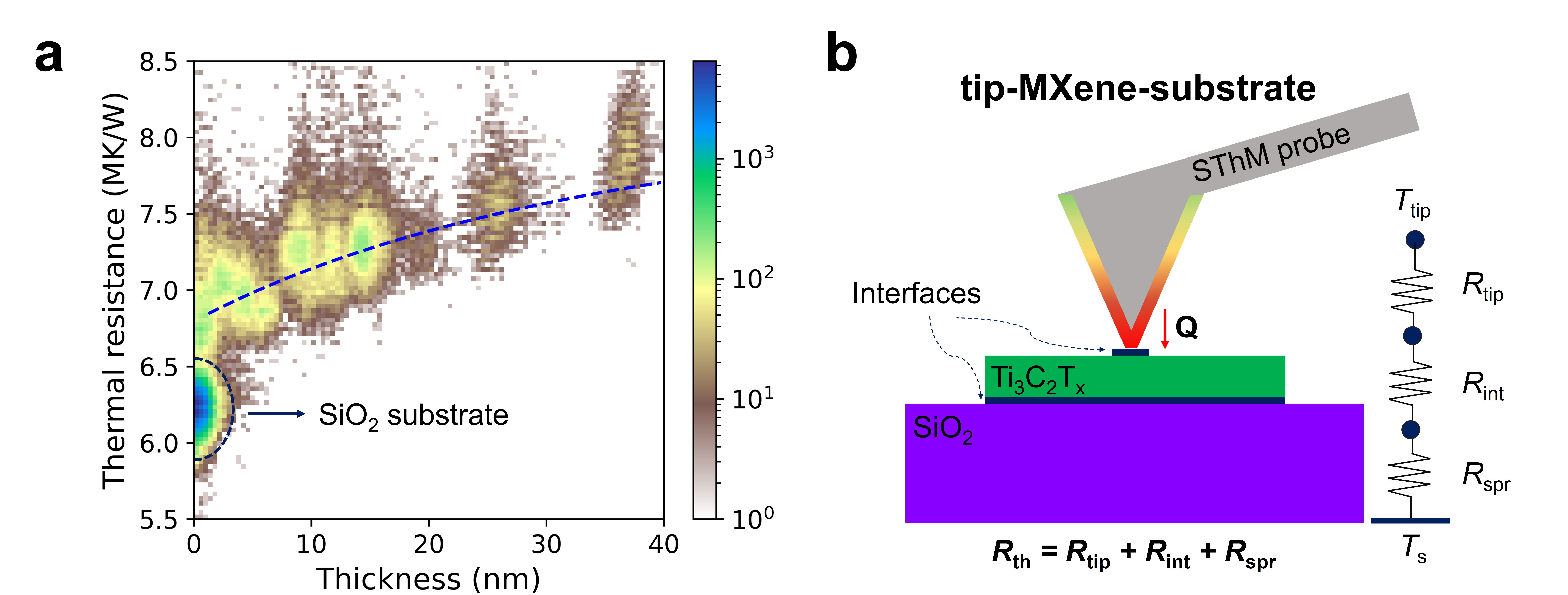} 
  \caption{(a) Total 2D histogram combined with all different Ti$_3$C$_2$T$_x$ thicknesses and the final fit curve (blue line) through a diffusive thermal transport model to extract thermal properties of Ti$_3$C$_2$T$_x$. (b) Schematic of the nanoscopic tip-sample contact and the heat transport model in the tip-MXene-substrate channel used in our experiments.} 
  \label{fig:example} 
\end{figure}

\begin{table}
\centering
  \caption{Extracted fitting parameters through a diffusive thermal transport model}
  \resizebox{\columnwidth}{!}{
  \begin{tabular}[htbp]{c|c|c|c|c|c}
    \hline
    Material  & $\kappa_\mathrm{i}$ [W m$^{-1}$ K$^{-1}$] & $\kappa_\mathrm{c}$ [W m$^{-1}$ K$^{-1}$] & $\kappa_\mathrm{eff}$ [W m$^{-1}$ K$^{-1}$] & $r_\mathrm{int}$ [K m$^2$ W$^{-1}$] & $R_\mathrm{tip}$ [K W$^{-1}$] \\
    \hline
    Ti$_3$C$_2$T$_x$/SiO$_2$  & $\qtyrange{0.85}{1.56}{}$ & $\qtyrange{0.38}{0.63}{}$ & 0.78 $\pm$ 0.21 & $1.0 \times 10^{-8}$ & $3.54 \times 10^{6}$ \\
    \hline
  \end{tabular}}
\end{table}

In the following, we quantify the thermal conductivity of Ti$_3$C$_2$T$_x$ using the data in Figure 3a. We assume a heat transfer mechanism in the tip-MXene substrate channel, which is depicted in Figure 3b. The total thermal resistance measured $R_\mathrm{th}$ is composed of a series of three resistances $R_\mathrm{th}$ = $R_\mathrm{tip}$ + $R_\mathrm{int}$ + $R_\mathrm{spr}$, where $R_\mathrm{tip}$ is the thermal resistance of the tip, $R_\mathrm{int}$ is the tip-sample interface resistance, and $R_\mathrm{spr}$ is the thermal spreading resistance. We further assume diffusive heat transport for orthotropic thermal spreading in layered Ti$_3$C$_2$T$_x$ flakes, according to previous thermal transport studies on other 2D materials \cite{luo2015anisotropic, kim2021extremely}. The orthotropic system has thermal conductivities dependent on the direction in the plane ($\kappa_\mathrm{i}$) and the direction between planes ($\kappa_\mathrm{c}$). An analytical expression for $R_\mathrm{spr}$ \cite{spiece2022low,evangeli2019nanoscale,gonzalez2023direct} was applied to fit the thermal resistance as a function of the thickness of the flake. The resulting fit is shown in Figure 3a as a blue dashed curve, and the fitting parameters are summarized in Table 1. Based on this thermal transport model, we find an interfacial thermal resistance between Ti$_3$C$_2$T$_x$ and SiO$_2$ of $r_\mathrm{int}$ of $1.0 \times 10^{-8}$ K m$^2$ W$^{-1}$, which is lower than the reported values of interface MoS$_2$/SiO$_2$ \cite{yasaei2017interfacial}, and close to that of interface graphene/SiO$_2$ \cite{chen2009thermal}. Such a low $r_\mathrm{int}$ greatly improves the sensitivity and reliability of our SThM measurements \cite{pernot2021frequency}. Furthermore, we obtain $\kappa_\mathrm{i}$ = $\qtyrange{0.85}{1.56}{}$ W m$^{-1}$ K$^{-1}$ and $\kappa_\mathrm{c}$ = $\qtyrange{0.38}{0.63}{}$ W m$^{-1}$ K$^{-1}$, respectively. Thus, the effective thermal conductivity $\kappa_\mathrm{eff}$ = 0.78 $\pm$ 0.21 W m$^{-1}$ K$^{-1}$ is lower than the $\kappa$ = 1.4 W m$^{-1}$ K$^{-1}$ of SiO$_2$, which is consistent with the contrast observed on the SThM maps (see, e.g., Figure 2c).

\begin{figure}[htbp] 
  \centering
  \includegraphics[width=15cm]{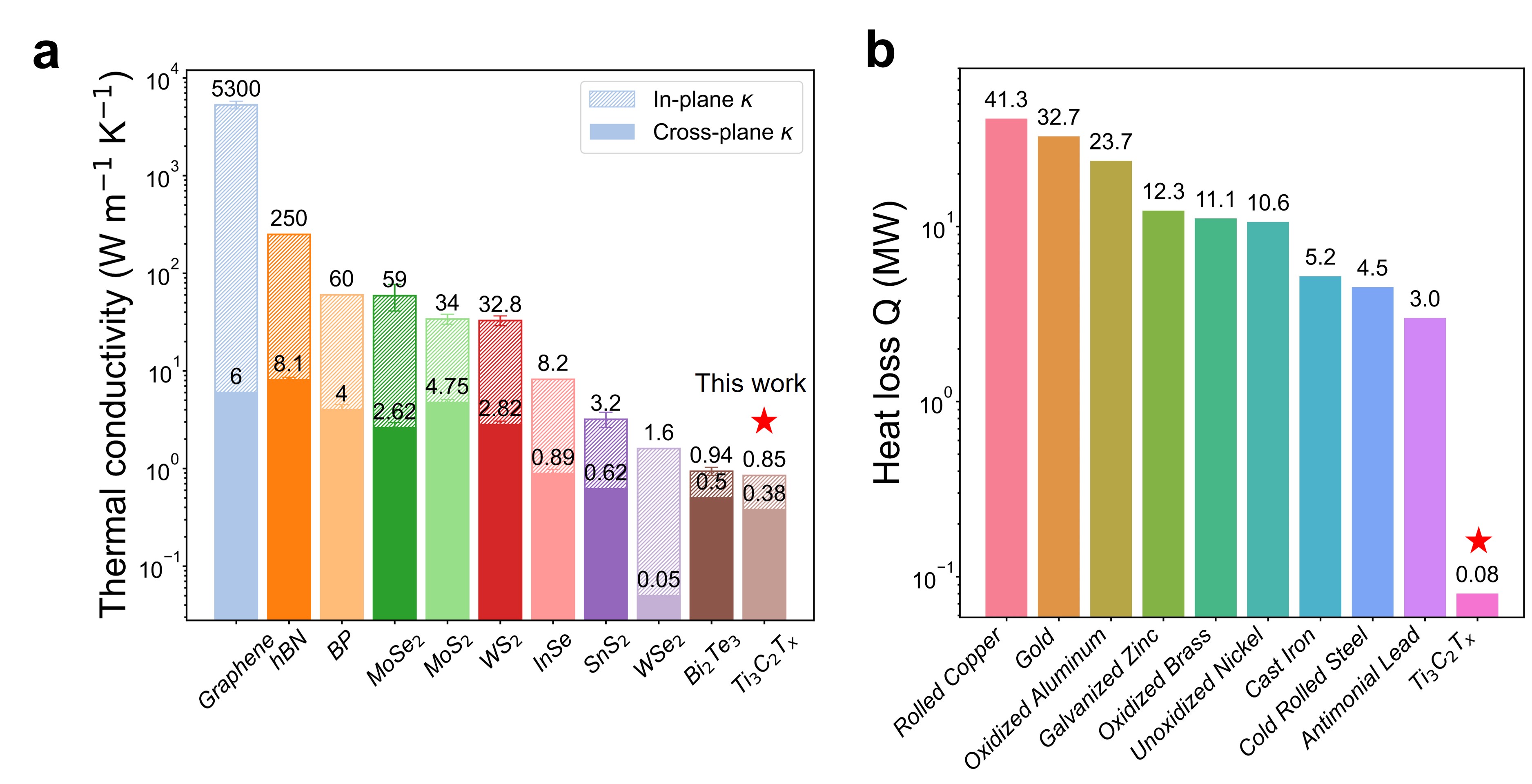} 
  \caption{(a) Comparison of experimental results in thermal conductivity for several 2D materials with monolayers and few layers, including graphene \cite{balandin2008superior(graphene), pop2012thermal(graphene-kc)}, hBN \cite{jo2013thermal(hbn), jaffe2023thickness(hbn-kc)}, BP \cite{jang2015anisotropic(bp)}, MoSe$_2$ \cite{zhang2015measurement(mose2), jiang2017probing(mos2ws2mose2-kc)}, MoS$_2$ \cite{dobusch2017thermal(mos2), jiang2017probing(mos2ws2mose2-kc)}, WS$_2$ \cite{sang2022measurement(ws2), jiang2017probing(mos2ws2mose2-kc)}, InSe \cite{buckley2021anomalous(inse), gonzalez2023direct}, SnS$_2$ \cite{karak2021low(sns2), bai2023interlayer(sns2-kc)}, WSe$_2$ \cite{chiritescu2007ultralow, mavrokefalos2007plane(wse2)}, Bi$_2$Te$_3$ \cite{pettes2013effects(bi2te3), teweldebrhan2010exfoliation(bi2te3-kc)}, and our work on Ti$_3$C$_2$T$_x$. (b) Comparison of the heat loss for Ti$_3$C$_2$T$_x$ MXene and other common metals in thermal management applications (see Note S5 in SI for detailed calculations).} 
  \label{fig:example} 
\end{figure}

We further compared the resulting in-/cross-plane thermal conductivities of Ti$_3$C$_2$T$_x$ with other single-crystalline 2D materials with similar thickness, including isolated mono- or few-layer flakes and thin films (Figure 4a). Compared to other 2D materials, Ti$_3$C$_2$T$_x$ shows a record low thermal conductivity, especially in in-plane direction. Surprisingly, this thermal conductivity is almost one order of magnitude smaller than the value estimated using the WF law, where contributions from phonons are neglected: by using the electrical resistance 3.18 k$\Omega$ measured in Figure S5 (See SI), we can estimate an electrical conductivity of $\sigma$ = 4.43 $\times 10^5$ S m$^{-1}$ at room temperature (See Note S4 in SI) \cite{lipatov2021high}, which is comparable to literature values reported for Ti$_3$C$_2$T$_x$ single flakes\cite{lipatov2024metallic}. Here, we used the entire thickness of the flake (measured by AFM) instead of the theoretical value (0.98 \si{\nano\metre}) in electrical measurement. This is the most conservative estimate possible, and still, we observe the breakdown of the WF law. Using Equation \ref{eq1} we can estimate $\kappa_\mathrm{WF} = $~3.17 W m$^{-1}$ K$^{-1}$, and thus an effective Lorenz number of $L$ = 0.25$L_0$. This strong violation of the WF law will be discussed in the following section. The low thermal conductivity found in Ti$_3$C$_2$T$_x$ makes the material attractive for thermal insulation or infrared stealth applications, and we compared its thermal insulation performance (heat loss) with other common metals in thermal management applications (Figure 4b). The heat loss of Ti$_3$C$_2$T$_x$ is two orders of magnitude smaller than common metals, such as gold, aluminium and steel, which shows great potential for use as a good thermal barrier in electronic devices and a large variety of other applications. Replacing polished metal casing of industrial equipment with Ti$_3$C$_2$T$_x$ foil or use of a submicrometer MXene coating on equipment may save billions of dollars in heat losses.

\subsection{Discussion}
The thermal conductivity of a material contains contributions from electrons ($\kappa_\mathrm{e}$) and phonons ($\kappa_\mathrm{ph}$): $\kappa_\mathrm{eff}$ = $\kappa_\mathrm{e}$ + $\kappa_\mathrm{ph}$. Since Ti$_3$C$_2$T$_x$ displays a metallic nature with a high density of electronic states near the Fermi level, we can assume that the total thermal conductivity is dominated by electron contributions ($\kappa_\mathrm{e}$ = $\kappa_\mathrm{eff}$ = 0.78 W m$^{-1}$ K$^{-1}$), and that the WF should therefore be valid. However, the observed $L$ = 0.25$L_0$ suggests a strong violation of the WF law. The WF law is a consequence of Fermi liquid theory and has been verified in numerous metals, such as gold or copper, where transport can be successfully described by weakly interacting Landau quasiparticles. In recent years, several violations of the WF law have been reported in strongly correlated systems with strong inelastic scattering of quasiparticles, where the Fermi liquid theory breaks down. Such systems involve graphene at the charge neutrality point \cite{crossno2016observation}, or some transition metal compounds \cite{lee2017anomalously}.

Usually, IV- and V-group MXenes (especially Ti$_3$C$_2$T$_x$) are metallic with high free carrier densities ($\approx$ 10$^{22}$ cm$^{-3}$), which drastically enhances interactions between charge carriers and phonons. This was confirmed in recent experimental studies \cite{zhang2022simultaneous, guzelturk2023understanding, colin2023ultrafast} that have provided evidence of strong correlations in MXenes. Using ultra-fast spectroscopy techniques, an electron-phonon coupling strength two orders of magnitude greater than that typically found for conventional metal counterparts was measured in Ti$_3$C$_2$T$_x$. Such strong electron-phonon coupling was further observed theoretically and experimentally in another MXene, Nb$_2$C  \cite{huang2019abnormally}. Furthermore, high photothermal conversion efficiencies \cite{wang2019mxene} and high phonon frequencies \cite{zhang2022simultaneous} were found in Ti$_3$C$_2$T$_x$, which also indicate strong electron-phonon coupling.

These strong correlations found in Ti$_3$C$_2$T$_x$ might \replaced{reveal}{shed light on} the mechanism responsible for the violation of the WF law observed in our experiments. Electron-phonon coupling is mainly induced by transverse optical (TO) and longitudinal optical (LO) phonon modes via the short-range deformation potential and the long-range Fröhlich interaction, respectively. Previous studies \cite{zhang2022simultaneous, zheng2022band} indicate strong electron-TO coupling but weak electron-LO coupling in Ti$_3$C$_2$T$_x$ with a higher density of states of TO modes. Energetic electrons strongly interact with TO phonons, which introduces an additional energy dissipation pathway. The weak electron-LO coupling suggested the formation of large polarons based on the Fröhlich polaron theory, which increases the effective mass and decreases charge mobility. These behaviors could be attributed to the reduction of $\kappa_\mathrm{e}$ and the breakdown of the WF law. In addition, $\kappa_\mathrm{e}$ and $\kappa_\mathrm{ph}$ may potentially be coupled because of the strong electron-phonon interaction. A theoretical calculation \cite{gholivand2019effect} predicted the $\kappa_\mathrm{ph}$ of Ti$_3$C$_2$T$_x$ ranging from 11 W m$^{-1}$ K$^{-1}$ (with the oxygen surface group) to 108 W m$^{-1}$ K$^{-1}$ (with the fluorine surface group). In our experiment, we expected a mixture of three kinds of surface terminations (-F, -OH, =O). The measured thermal conductivity is still much lower than the calculated values, which consider phonon-phonon and boundary scatterings. This suggested that local defects and inelastic electron-phonon scattering processes may result in a large limitation of the phonon heat transport and a further reduction of $\kappa_\mathrm{ph}$. Therefore, we conclude that the ultralow thermal conductivity of Ti$_3$C$_2$T$_x$ single flakes is mainly caused by strong electron-phonon interactions.

\section{Conclusion}
We experimentally quantified the anisotropic thermal transport properties of isolated single-crystalline Ti$_3$C$_2$T$_x$ flakes by combining scanning thermal microscopy at room temperature and a diffusion heat transfer model. An ultralow thermal conductivity of $\kappa_\mathrm{eff}$ = 0.78 $\pm$ 0.21 W m$^{-1}$ K$^{-1}$ was observed. Given the high electrical conductivity of $\sigma$ = 4.43 $\times 10^5$ S m$^{-1}$ found in our samples, we reveal a strong violation of the WF law, with an effective Lorenz number $L$ = 0.25$L_0$. We attribute this violation to the strong electron-phonon coupling in Ti$_3$C$_2$T$_x$, especially the electron-LO scattering combined with local defects or phonon scatterings that reduce the phononic thermal conductivity. These results provide an experimental basis showing that MXenes are very promising materials for future thermoelectric applications thanks to a high electrical conductivity combined with high thermal resistivity. Furthermore, this work highlights the application potential of Ti$_3$C$_2$T$_x$ for thermal management or -- thanks to its very low infrared emissivity \cite{han2023versatility} -- for thermal barrier coatings.

\section{Experimental Section}

\subsection{Preparation of Ti$_3$C$_2$T$_x$ colloid}
Ti$_3$C$_2$T$_x$ was etched from Carbon Ukraine Ti$_3$AlC$_2$ MAX phase. The etchant was prepared by mixing 6 mL deionized water, 2 mL hydrofluoric acid (51 wt.\% aq. Thermo Fisher Scientific), and 12 mL hydrochloric acid (38 wt.\% aq. Thermo Fisher Scientific). 1 g Ti$_3$AlC$_2$ was then etched for 24 hours at 35 \si{\celsius} and 300 rpm for the stir bar. After etching out the Al, the acid was washed out of the multilayer Ti$_3$C$_2$T$_x$ through 8 cycles of centrifugation at 3500 rpm for 3 min per cycle. Once the solution was pH neutral, H$_2$O was added to make a 20 mL solution, and 0.424 g LiCl was added to make a 0.5 M concentration of Li$^+$ ions, which were then intercalated into the Ti$_3$C$_2$T$_x$ by mixing for 24 hours at 35 \si{\celsius} and 300 rpm. After intercalation, the monolayer and few-layer MXene flakes were isolated through repeated centrifugation and skimming, as only the single flakes remained suspended for collection after centrifugation. Centrifugation conditions were 10 min and 3500 rpm to remove LiCl and repeated until supernatant was no longer clear, with subsequent increases in cycle time for single-layer MXene supernatant collection. About 10-12 cycles were performed, and then the predominantly single-layer Ti$_3$C$_2$T$_x$ dispersion was concentrated through centrifugation at 10000 rpm for 20 min.

\subsection{Fabrication of Ti$_3$C$_2$T$_x$ devices}
Samples of Ti$_3$C$_2$T$_x$ isolated flakes were obtained by drop-casting MXene aqueous dispersion onto a Si/SiO$_2$ (290 nm-thick SiO$_2$) substrate, followed by washing with slowly flowing deionized water and naturally dried under a flow of nitrogen gas. The wafers were finally put in an oven at 60 \si{\celsius} to completely dry the flakes. Before deposition, the Si/SiO$_2$ wafer was ultrasonically cleaned in acetone and then deionized water for 10 min to ensure the cleanliness of the surface. We obtained lots of single MXene flakes with different thicknesses within the drop-cast area. The indium needles were fabricated in Ti$_3$C$_2$T$_x$ single flake as electrical contact, which can achieve good ohmic contacts. In this technique \cite{girit2007soldering}, the substrate with indium bead was first heated to around 165 \si{\celsius} and the indium melted. Then, a tungsten tip was inserted into the melting pool and slowly pulled out to form a sharp needle, which was finally transferred to MXene flakes to form the contacts.

\subsection{SThM measurements on Ti$_3$C$_2$T$_x$ devices}
Measurements were performed by a Bruke Dimension Icon scanning probe microscope platform with a commercial SThM probe (KNT-SThM-2an, Kelvin Nanotechnology) under an ambient environment. The SThM probe has a resistive palladium heater with a resistor of $\approx$ 350 Ohms at room temperature and a tip radius of $\approx$ 75 nm (Figure 2d). Similar to our previous work \cite{evangeli2019nanoscale}, we first zeroed the Wheatstone bridge and heated the probe by applying an AC voltage of 2 V at a frequency of 91 kHz with a 2 V DC offset. The probe was scanned over the sample area with contact mode, and the thermal response $V_\mathrm{SThM}$ was measured with lock-in amplifier (SRS830) pixel by pixel from the unbalanced bridge caused by the local thermal conductance changes of the sample. By monitoring $V_\mathrm{SThM}$, we can create a thermal resistance map of the MXene sample. Finally, the amplified signals were recorded by scanning probe microscope controller for data acquisition using commercial software. Topography and thermal maps of Ti$_3$C$_2$T$_x$ devices could be obtained simultaneously.

\subsection{Diffusive thermal transport model}
MXenes are anisotropic materials with directional-dependent thermal conductivities. To transform such an orthotropic system to an effective isotropic thermal conductivity, we consider an effective thermal conductivity defined as $\kappa_\mathrm{eff} = \sqrt{\kappa_\mathrm{i} \times \kappa_\mathrm{c} }$ and an effective thickness defined as $t_\mathrm{eff} = t\sqrt{\frac{\kappa_\mathrm{i} }{\kappa_\mathrm{c} } } + r_\mathrm{int} \kappa_\mathrm{eff}$ (t is the physical thickness, $r_\mathrm{int}$ is the thermal interface resistivity). Here, the first term accounts for the anisotropy and the second includes the MXene-SiO$_2$ interface thermal resistance. Muzychka \cite{muzychka2004thermal} and Spiece \cite{spiece2022low} derived an analytical expression for the thermal spreading resistance: 
\begin{equation}
R_\mathrm{spr}(t) = \frac{1}{\pi \rho \kappa_\mathrm{eff} } \int\limits_{0}^{\infty} \left [ \frac{1+Ke^{(\frac{-2\xi t_\mathrm{eff} }{\rho })} }{1-Ke^{(\frac{-2\xi t_\mathrm{eff} }{\rho })}}  \right ]J_{1}(\xi ) \sin (\xi)\frac{\mathrm{d}\xi}{\xi ^{2} } 
\end{equation}
Where $J_1$ is the Bessel function of the first kind of order, $\rho$ is the tip radius, $\xi$ the integration variable, and K is defined as $K = (1-\frac{\kappa_\mathrm{sub} }{\kappa_\mathrm{eff} }) /(1+\frac{\kappa_\mathrm{sub} }{\kappa_\mathrm{eff}})$.

For very thin flakes (less than 10 nm), because the radius of the tip (75 nm) is much larger than flake thickness, the heat dissipation channel from the tip to the substrate is almost vertical (cross-plane direction). With this assumption, we can first use an isotropic thermal model (let $\kappa_\mathrm{i}$ = $\kappa_\mathrm{c}$) to estimate the cross-plane thermal conductivity $\kappa_\mathrm{c}$. We used this unknown $\kappa_\mathrm{c}$ to fit Equation 2 with experimental data (thickness-dependent thermal resistance). In detail, we combined seven 2D histogram maps into one and divided it into multiple small bins. For each column, we selected the bin containing the most data points. The average value of the data in each bin was then used for fitting. When it reached the best fit, we extracted the value of $\kappa_\mathrm{c}$. Then we used this $\kappa_\mathrm{c}$ value to fit thicker flakes, we considered the orthotropic model to calculate the total thermal resistance $R_\mathrm{th}$ = $R_\mathrm{tip}$ + $R_\mathrm{int}$ + $R_\mathrm{spr}$ and used the unknown parameters $\kappa_\mathrm{i}$, $r_\mathrm{int}$ and $R_\mathrm{tip}$ to fit the experimental data $R_\mathrm{th}$. As a result, we obtained the $\kappa_\mathrm{i}$, $\kappa_\mathrm{c}$, $r_\mathrm{int}$ and $R_\mathrm{tip}$ values.

\begin{acknowledgement}

The authors acknowledge financial support from the F.R.S.-FNRS of Belgium (FNRS-CQ-1.C044.21-SMARD, FNRS-CDR-J.0068.21-SMARD, FNRS-MIS-F.4523.22-TopoBrain, FNRS-CR-1.B.463.22-MouleFrits, FNRS-PDR-T.0029.22-Moire), from the Federation Wallonie-Bruxelles through the ARC Grant No. 21/26-116 and from the EU (ERC-StG-10104144-MOUNTAIN). This project (40007563-CONNECT) has received funding from the FWO and F.R.S.-FNRS under the Excellence of Science (EOS) programme. Y.H. acknowledges support from the China Scholarship Council and Wallon-Brussels International (CSC-WBI funding, project No. 202108440051). Research at Drexel University was supported by U.S. National Science Foundation under Grant CHE-2318105 (M-STAR CCI).

\end{acknowledgement}

\begin{suppinfo}

The following files are available free of charge.
\begin{itemize}
  \item Raman spectrum of the monolayer Ti$_3$C$_2$T$_x$; SThM probe calibration (Probe approach and retraction curves, probe temperature as a function of the supplied power); SEM images of the SThM probe; SThM thermal map on a few layers Ti$_3$C$_2$T$_x$; I-V curve of a two-terminal Ti$_3$C$_2$T$_x$ device; Thermal management applications of Ti$_3$C$_2$T$_x$ (Heat loss model and calculations) (PDF).
\end{itemize}

\end{suppinfo}

\bibliography{references}

\end{document}